\documentclass[preprint,10pt]{elsarticle}
\usepackage{epsfig}          
\usepackage{latexsym}
\usepackage{amssymb}
\usepackage{amsmath}

\usepackage{ifthen}
\usepackage{color}
\def\vls{yes}
\newcommand{\novls}{\def\vls{no}}
\newcommand{\vl}[1]{\ifthenelse{\equal{\vls}{no}}{}{
  \marginpar{\vspace{0.7cm}\hspace{0.5cm}\parbox{2.3cm}
{\small  \textcolor{red}{\em{ #1}}}}}}

\def \vlabels {yes}
\newcommand{\novlabels}{\def \vlabels {no}}
\newcommand{\vlabel}[1]{\ifthenelse{\equal{\vlabels}{no}}{\label{#1}}{\label{#1} \tag{\textcolor{red}{#1}}}}

\def \todos {yes}
\newcommand{\notodos}{\def \todos {no}}
\newcommand{\todo}[1]{\ifthenelse{\equal{\todos}{no}}{}{
  \marginpar{{\small  \textcolor{blue}{\em{ #1}}}}}}

\def\bems{yes}

\newcommand{\bem}[1]{\ifthenelse{\equal{\bems}{no}}{}
{\small \textcolor{green}{ \em{ #1}}}}

\begin{document}

\novlabels
\novls
\notodos

\begin{frontmatter}

%% Title, authors and addresses

%% use the tnoteref command within \title for footnotes;
%% use the tnotetext command for the associated footnote;
%% use the fnref command within \author or \address for footnotes;
%% use the fntext command for the associated footnote;
%% use the corref command within \author for corresponding author footnotes;
%% use the cortext command for the associated footnote;
%% use the ead command for the email address,
%% and the form \ead[url] for the home page:
%%
%% \title{Title\tnoteref{label1}}
%% \tnotetext[label1]{}
%% \author{Name\corref{cor1}\fnref{label2}}
%% \ead{email address}
%% \ead[url]{home page}
%% \fntext[label2]{}
%% \cortext[cor1]{}
%% \address{Address\fnref{label3}}
%% \fntext[label3]{}

\title{Nonlinear conductance and noise in boundary sine-Gordon and related models}
%\tnotetext[star]{}
%% use optional labels to link authors explicitly to addresses:
%% \author[label1,label2]{<author name>}
%% \address[label1]{<address>}
%% \address[label2]{<address>}

\author[ITP3]{Jens Honer}
\ead{honer@itp3.uni-stuttgart.de}

\author[ITP2]{Ulrich Weiss\corref{cor1}}
\ead{weiss@theo2.physik.uni-stuttgart.de}

\cortext[cor1]{Corresponding author.}

\address[ITP3]{ 3. Institut f\"ur Theoretische Physik, Universit\"at Stuttgart, Pfaffenwaldring 57, D-70550,Stuttgart, Germany}
\address[ITP2]{ 2. Institut f\"ur Theoretische Physik, Universit\"at Stuttgart, Pfaffenwaldring 57, D-70550,Stuttgart, Germany}

\begin{abstract}
We study a conjecture by Fendley, Ludwig and Saleur for the nonlinear conductance in the boundary  sine-Gordon model. They have calculated the perturbative series of twisted partition functions, which require particular (unphysical) imaginary values of the bias, by applying the tools of Jack symmetric functions to the "log-sine" Coulomb gas on a circle. We have analyzed the conjectured relation between the
analytically continued free energy and the nonlinear conductance in various limits. We confirm the conjecture for  weak and
strong tunneling, in the classical regime, and in the zero temperature limit. We also shed light on this special variant of the 
${\rm Im}\,F$-method and compare it with the real-time Keldysh approach. In addition, we address the issue of quantum statistical
fluctuations.

\end{abstract}

\begin{keyword}
%% keywords here, in the form: keyword \sep keyword
Quantum transport \sep full counting statistics \sep quantum impurity problem \sep Coulomb gas \sep Jack polynomials

%% PACS codes here, in the form: \PACS code \sep code
\PACS 05.60.Gg \sep 05.40.Ca \sep 71.10.Pm \sep 73.50.Td  
%% MSC codes here, in the form: \MSC code \sep code
%% or \MSC[2008] code \sep code (2000 is the default)
\end{keyword}

\end{frontmatter}

%%
%% Start line numbering here if you want
%%
% \linenumbers

%-----------------------------------------------------------------------------------------------------------------------

%% main text
\section{Introduction}

Quantum Brownian motion (QBM) of a particle or charge in a tilted periodic potential is one of the basic subjects of 
nonequilibrium quantum-statistical mechanics and a key model for a multitude of quantum transport phenomena in condensed matter
\cite{qds08}.
Of particular interest is a duality symmetry between the weak- and strong-binding representations of the  
model \cite{Schmid,FishZwerg}. There is a close correspondence of the QBM model with seemingly different models: (i)  a resistively
shunted  Josephson junction  \cite{sz90}, (ii)  a quantum impurities in a 1D wire \cite{kf92}, (iii) tunneling of edge currents through constrictions in fractional quantum-Hall 
devices \cite{wen90}, and (iv) a coherent one-channel conductor in a resistive electromagnetic environment \cite{SafiSaleur}. 
In fact, all these models are directly related to each other. 

The calculation of the dynamics may be done with the use of the Feynman-Vernon influence functional method \cite{FeynVern} or the 
equivalent nonequilibrium Keldysh technique. In this theory, the reduced density matrix (RDM) is expressed as a double path sum
over all possible paths on the RDM-plane. The path sum problem is formally equivalent to the grand-canonical sum of a charge gas with 
complex-valued interactions. Of particular interest is the scaling or field theory limit. Until now the respective 
real-time Coulomb gas has not been solved analytically barring special cases. 

One may also consider the system in imaginary time compactified on a circle of length $1/T$, where $T$ is the temperature. In the
field theory limit, the charges of the respective Coulomb gas interact with a "log-sine" long range interaction.  The
compactified Coulomb gas has been solved analytically for particular unphysical values of the bias with the tool of Jack symmetric functions \cite{fls95jsp}. The consideration of nonequilibrium quantum transport requires (i) analytic 
continuation of the calculated partition function to the physical bias, and (ii) a conjectured relation between the 
analytically continued partition function and transport quantities \cite{fls96jsp}. Here we analyze this route
in different parameter regimes, and we compare the findings with results available by different methods.

Our paper is organized as follows. In Section~\ref{sgtm} we introduce several related models, which may describe nonequilibrium 
quantum transport of a particle or a charge, Josephson junction dynamics, scattering off an impurity embedded in a Luttinger liquid,
and conductance of one-channel conductors in a resistive electromagnetic environment.
In Section \ref{scgr} we review the solution of the "log-sine" Coulomb gas on a circle with winding numbers.
In Section \ref{scond} we study the conjecture for the nonlinear conductance in the weak-tunneling regime, in the classical regime,
and in the zero temperature limit. In Section \ref{snoise} we compare the results of the imaginary time approach with those of
the real-time Keldysh method. We also address the issue of statistical fluctuations. 
Section~\ref{sconcl} is devoted to concluding remarks.

\section{Generic transport model}
\label{sgtm}

The boundary sine-Gordon (BSG) model is a one-dimensional quantum field with gapless bulk excitations and a sinusoidal  boundary interaction. The model is integrable and can be treated by a variety of powerful techniques. It constitutes a generic quantum transport model which displays a wide range of interesting characteristics, and it is a realistic model to describe seemingly different physical situations.
% (i) a quantum Brownian particle moving in a tilted washboard potential, 
% (ii) charge transport through a quantum impurity embedded in a 1D conductor, (iii) a coherent one-channel 
% conductor in a resistive electromagnetic environment, and (iv) a resistively shunted Josephson junction. 
In fact, all these models are closely related to each other. They have in common that a nonlinear quantum degree of freedom
is coupled to a harmonic field.

\subsection{Nonlinear quantum Brownian motion}

A quantum Brownian particle moving in a periodic potential is a 
key model for many transport phenomena in condensed matter. 
Of particular interest is the case of strict Ohmic dissipation in which the regions of weak and
strong corrugation are related by an exact self-duality. 
\subsection{The dual representations}
\label{sbrowndual}
The dissipative dynamics of a particle moving in a tilted, sinusoidally corrugated potential of periodicity length $X_0^{}$ may be described by the translational-invariant
Hamiltonian with bilinear environmental coupling\footnote{Throughout this paper we put $\hbar = k_{\rm B}^{} =1$. }

\begin{equation}\label{hwb}
\begin{array}{rcl}
	H_{\rm WB}^{} &=& {\displaystyle  \frac{P^2}{2M}-U_0\cos\bigg(\,\,\frac{2\pi X}{X_0^{}}\bigg)\,-\, 
\epsilon_{\rm WB}^{} \frac{X}{X_0^{}} }\\[4mm]
&& \!\!\!\!\!+ \, {\displaystyle
\sum_\alpha \omega_\alpha\left(b_\alpha^\dagger- \frac{\lambda_\alpha}{\omega_\alpha}\frac{X}{X_0^{}}\right)
\left(b_\alpha^{}- \frac{\lambda_\alpha}{\omega_\alpha}\frac{X}{X_0^{}}\right)         \;.                 }
\end{array}
\end{equation}
The "weak-binding" form (\ref{hwb}) is the convenient representation for the perturbative expansion of transport
quantities in the corrugation strength $U_0^{}$. Of particular interest is the normalized mobility $\mathcal{M} =\mu/\mu_0$ (particle transport)
or normalized conductance $\mathcal{G} =G/G_0$ (charge transport), where $\mu_0$ and $G_0$ are mobility and conductance in the absence of the corrugation. We have $\mathcal{M}=\mathcal{G}$. In the sequel we use the language of charge transport.

In the tight-binding regime $T/\omega_0^{},\; \epsilon_{\rm TB}^{} /\omega_0^{}\;\ll\; 1\; \ll\;U_0/\omega_0^{}$,
where $\omega_0^{}$ is the distance of low-lying levels in the individual well regimes, and $\epsilon_{\rm TB}^{}$ is
the potential drop between neighboring wells, only the lowest energy band is occupied. Then the system is effectively reduced
to a tight-binding lattice. The tight-binding or tunneling Hamiltonian  with transfer matrix element $\Delta$ and translational 
invariant coupling is  
\begin{equation}\label{htb}
\begin{array}{rcl}
 H_{\rm TB}^{} &=& -\,{\displaystyle  \frac{\Delta}{2} 
 \sum_n \Big(\, |n\rangle\langle n+1| \,+\, |n+1\rangle\langle n| \, \Big)   \,- \,  \epsilon_{\rm TB}^{}\frac{Y}{Y_0^{}} }  \\[4mm]
 && \!\!\!\!\! + \,{\displaystyle  
 \sum_\alpha \Omega_\alpha\left(b_\alpha^\dagger- \frac{\mu_\alpha}{\Omega_\alpha}\frac{Y}{Y_0}\right)
\left(b_\alpha^{}- \frac{\mu_\alpha}{\Omega_\alpha}\frac{Y}{Y_0}\right)  }  \; ,
\end{array}
\end{equation}
where $Y_0^{}$ is the TB lattice constant, and 
\begin{equation} 
Y \;=\;   Y_0^{} \sum_n n\, |n\rangle\langle n| 
\end{equation}
is the position in TB representation. Since we postulate the same tilting force in both models, the potential drops
between neighboring sites are related by
\begin{equation}
\epsilon_{\rm WB}^{}/X_0^{} \;=\; \epsilon_{\rm TB}^{}/Y_0^{} \; .
\end{equation}

In the scaling or field theory limit, the spectral densities of the environmental couplings,
\begin{equation}\label{densspec}
\begin{array}{rcl}
 G_{\rm WB}^{}(\omega) &=& \sum_{\alpha}
\lambda_\alpha^2 \, \delta(\omega-\omega_{\alpha})\; , \\[2mm]
G_{\rm TB}^{}(\omega) &=&  \sum_\alpha\mu_\alpha^2 \,\delta(\omega-\Omega_{\alpha}) \; ,
\end{array}
\end{equation}
are of strict Ohmic form
\begin{equation}\label{gft}
\begin{array}{rcl}
G_{\rm WB}(\omega) &=& 2 K_{\rm WB}\, \omega \;, \\[2mm] 
G_{\rm TB}(\omega) &=& 2 K_{\rm TB}\, \omega  \; .
\end{array}
\end{equation}
The dimensionless so-called Kondo parameters $K_{\rm WB}$ and $K_{\rm TB}$ are related to the Ohmic viscosity $\eta$ by 
\begin{equation}\label{Krel} 
K_{\rm WB}^{}\;=\; \frac{\eta X_0^2}{2\pi}\;, \qquad K_{\rm TB}^{}\;=\; \frac{\eta Y_0^2}{2\pi}\;. 
\end{equation}
In the Ohmic regime (\ref{gft}), the TB and WB model (\ref{hwb}) and (\ref{htb}) are related by an exact self-dual symmetry, 
as noted first by Schmid for the equilibrium density matrix at $T=0$ \cite{Schmid}. Later,
the duality was proven to hold also at finite $T$ and for real time dynamics \cite{FishZwerg}. In the self-dual mapping we have 
\begin{equation} \label{dual1}
Y_0^{}/X_0^{} \;=\; K_{\rm TB}\equiv K\;=\; 1/K_{\rm WB}^{} \; ,
\end{equation}
and the normalized conductance in the WB model $(\ref{hwb})$ is directly related to the corresponding quantity in the associated TB model
(\ref{htb}),
\begin{equation}\label{dual2}
\mathcal{G}_{\rm WB}^{} (T,\epsilon,K ) \;=\; 1- \mathcal{G}_{\rm TB}^{}(T,\epsilon/K,1/K )  \; . 
\end{equation}

The self-duality relation (\ref{dual2}) exchanges weak and strong corrugation, as well as the regimes of weak and strong environmental coupling.
The explicit calculation of (\ref{dual2}) may be executed with the use of the Feynman-Vernon influence functional method \cite{FeynVern}, 
or directly by unitary transformations of the WB and TB Hamiltonians \cite{Schom}.

Computations are conveniently performed in the discrete TB model. With use of the self-duality, the
results can then easily be transferred to the dual WB model (see below).

With regard to the partition function, we consider the system in imaginary time compactified on a circle of length $1/T$.
The partion functions $\mathcal{Z}_{\rm TB}$ and $\mathcal{Z}_{\rm WB}$ can be expanded in even powers of $\Delta$ and $U_0$, respectively. The perturbative series of the partition function is equivalent to the grand-canonical sum of a one-dimensional gas of positive and negative charges $\xi=\pm 1$ with overall neutrality. The charges represent the forward and backward moves in the TB or WB potential. 
All effects of the environmental coupling are described by the imaginary-time charge-charge correlator
\begin{equation}
\begin{array}{rcl}
\left\langle \mathcal{T}_\tau\,{\rm e}^{{\rm i}\,2\pi q(\tau)/q_0 }_{} \, 
              {\rm e}^{- {\rm i}\,2\pi q(0)/q_0 }_{}  \right\rangle  \!&=& \!
         {\rm e}^{-\,(2\pi/q_0)^2\,\langle\,[\,q(0)-q(\tau)\,]\,q(0)\,\rangle}_{} \\[4mm]   
            \! &=&  \!{\rm e}^{-\,W(\tau)}_{}  \; ,
\end{array}
\end{equation}
where $\langle\cdots\rangle$ means quantum statistical equilibrium average. The spectral representation of 
the charge interaction is
\begin{equation}\label{corrspec}
W(\tau) = \int_0^\infty  \!\! {\rm d}\omega \,\frac{G(\omega)}{\omega^2}\;\frac{\cosh(\frac{\omega}{2T}) - 
\cosh[\omega(\frac{1}{2T} -\tau))]}{\sinh(\frac{\omega}{2T})}    \;  .
\end{equation}
In the scaling limit (\ref{gft}) we have
\begin{equation}\label{corr1}
W(\tau) = 2 K \ln\left[\, \frac{\omega_{\rm c}^{}}{\pi T}\,\sin(\pi T \tau) \,\right] \;,
\end{equation}
where $\omega_{\rm c}^{}$ is the UV-frequency cutoff in the spectral density $G(\omega)$. 

While in a two-state system the charges alternate, in our case of infinitely many states the charges are unordered. 
In the term of order $\Delta^{2n}$ or $U_0^{2 n}$, we have $n$ positive and $n$ negative charges, and we must integrate over 
the locations of the charges in the interval $1/T$. The grand-canonical series of the TB partition function is
\begin{equation} \label{coul1}
\begin{array}{rcl}
\mathcal{Z}_{\rm TB} &=& {\displaystyle 1\,+\, \sum_{n\,=\,1}^\infty \frac{(\Delta/2)^{2n}}{(n!)^2} 
\int_0^{1/T}\!\! \prod_{i\,=\,1}^n{\rm d}\tau_i 
\int_0^{1/T}\!\! \prod_{j\,=\,1}^n{\rm d}\tau'_j  }\\[5mm]     
&&\!\times \;{\displaystyle \exp\left\{\sum_{m>\,k\,=\,1}^n\left[W(\tau_m^{}-\tau_k^{}) + W(\tau'_m-\tau'_k )\right] \right\}}  \\[5mm]  
&&\!\quad \times\; {\displaystyle  \exp\left\{-\sum_{k,\,m\,=\,1}^n W(|\tau_m^{}-\tau'_k|)       \right\}  } \\[5mm]    
&& \!\qquad\times \; {\displaystyle \exp\left\{-\;\epsilon\sum_{k\,=\,1}^n(\tau_k^{}-\tau'_k)\right\} } \; , 
\end{array}
\end{equation}
and a similar expression holds for $\mathcal{Z}_{\rm WB}$.

\subsection{Josephson junction dynamics}

According to the dc Josephson effect, a super current of Cooper pairs flows through a super\-conducting tunnel junction. 
At finite voltage $V_{\rm x}^{}$, there is a supercurrent only if the excess energy  
$ 2 e V_{\rm x}^{}$ can drain off to the electromagnetic environment. Quasiparticle excitations can be disregarded when 
the voltage $V_{\rm x}^{}$ is small compared to the gap voltage.
For weak Josephson coupling, $E_{\rm J}^{}\ll E_{\rm C}^{} = 2e^2_{}/C$, the charge on the junction is fairly well defined. The relevant minimal model of the junction-plus-environment entity is \cite{ingnaz92}
\begin{equation}  \label{jos2}
\begin{array}{rcl}
 H &=& {\displaystyle \frac{Q^2_{}}{2C} \, - \, E_{\rm J}^{}\cos \psi } \\[3mm]
&&\!\!\!\!\! +\;{\displaystyle  \sum_{\alpha\,=\,1}^N \Bigg[
\,\frac{q_\alpha^2}{2C_\alpha^{}} \,+ \, \frac{1}{e^2_{}}
\frac{1}{ 2 L_\alpha^{}}\left(\frac{\psi}{2} \,- \,eV_{\rm x}^{}t \,- \, \psi_\alpha^{}\right)^2\,\Bigg] }  \; . 
\end{array}
\end{equation}
The first term represents the charging energy, the second term accounts for tunneling of Cooper pairs through 
the junction, and the third term describes the coupling to the electromagnetic environment, which is modeled by a set of 
LC-circuits, and it also includes the applied voltage. 
In the limit $N\to \infty$, the capacitances $\{C_\alpha\}$ and inductances $\{L_\alpha\}$ provide an arbitrary linear admittance
\begin{equation}
Y(\omega)\;=\; Z^{-1}(\omega)\;=\; \int_0^\infty\!\!{\rm d}t\,{\rm e}^{-\,{\rm i}\,\omega t}_{}
\sum_{\alpha\,=\,1}^N\frac{\cos(\omega_\alpha^{} t)}{L_\alpha} \;, 
\end{equation}
where  $\omega_\alpha^{} = 1/ \sqrt{L_\alpha C_\alpha}$. The total impedance seen by the junction is
\begin{equation}
Z_{\rm t}^{}(\omega)\;=\; \frac{1}{ {\rm i}\,\omega C + Y(\omega)} \;. 
\end{equation}
After tracing out the electromagnetic modes, one finds for the imaginary-time phase correlator in thermal equilibrium
\begin{equation}
\left\langle \mathcal{T}_\tau\,{\rm e}^{{\rm i}\,\psi(\tau)}_{}\,{\rm e}^{-\,{\rm i}\,\psi(0)}_{}\right\rangle\;=\;
{\rm e}^{-\,\langle \,[\,\psi(0)-\psi(\tau)\,]\,\psi(0)\,\rangle}_{} \; \equiv \; {\rm e}^{-\,W_\psi(\tau)}_{} \; ,
\end{equation}
where $W_\psi(\tau)$ has the spectral representation  (\ref{corrspec}) with
\begin{equation}\label{gpsi}
G_\psi(\omega) \;=\; 2\, \frac{{\rm Re}\,Z_{\rm t}(\omega)}{R_{\rm Q}} \,\omega \;, 
\end{equation}
and where $R_{\rm Q}^{} = R_{\rm K}^{}/4 = \pi/2e^2_{} $ is the resistance quantum for Cooper pairs.
For a resistive electromagnetic environment, ${\rm Re}\,Z_{\rm t}(\omega)/R_{\rm Q} = 
\rho\, /[\,1+(\omega/\omega_{\rm R}^{})^2\,]$ with $\rho=R/R_{\rm Q}$, and $\omega_{\rm R}^{} = 1/[Z(0) C] = E_{\rm c}/(\pi\rho)$, 
we get 
\begin{equation}\label{wpsi}
W_\psi(\tau) \;=\; 2 \rho \ln\left[\, \frac{\omega_{\rm R}^{}}{\pi T}\,\sin(\pi T \tau) \,\right]  \,+ \, 2\rho\,\zeta  \; .
\end{equation}
The first term in Eq. (\ref{wpsi}) describes the resistive environment in the scaling limit [\,cf. Eq.~(\ref{corr1})\,], 
and the quantity $\zeta$ encapsulates details of the total impedance at high frequencies  \cite{ing94}, 
\begin{equation}\label{hifr}
\begin{array}{rcl}
\zeta &=& {\displaystyle \psi\left(1+ \frac{\omega_{\rm R}^{}}{2\pi T}\right) -\psi(1)-\ln\left(\frac{ \omega_{\rm R}^{}}{2\pi T} \right) }
 \\[3mm]
&&{\displaystyle \;+ \, \int_0^\infty \!\frac{{\rm d}\omega}{\omega}\left[\,\frac{ {\rm Re}\,Z_{\rm t}(\omega)}{\rho R_{\rm Q}} -
\frac{1}{ 1+(\omega/\omega_{\rm R}^{})^2}   \,\right]  }  \; .
\end{array}
\end{equation}

The second term in (\ref{wpsi}) leads to  adiabatic renormalization of the tunneling
coupling
\begin{equation}\label{dressed}
E_{\rm J}\quad\to\quad E'_{\rm J} \;=\; E_{\rm J} \, {\rm e}^{-\rho\,\zeta} \; .
\end{equation}

The model (\ref{jos2}) with (\ref{wpsi}) directly corresponds to the model
(\ref{htb}) with (\ref{gft}). The correspondence relations are
\begin{equation}\label{Jos}
\Delta \;\; \hat{=}\;\; E'_{\rm J}\; , \qquad\epsilon \;\;\hat{=}\;\;  2eV_{\rm x}^{} \; ,
 \qquad K_{}^{}\;\;\hat{=}\;\; \rho   \;  .
\end{equation}

In the opposite limit of large Josephson coupling, $E_{\rm J}\gg E_{\rm c}$, the phase of the BCS condensate is fairly well defined and localized in
one of the wells of the Josephson potential. Then it is pertinent to turn to the Bloch band representation of the Josephson junction and to
choose a basis of discrete phase states,
\begin{equation}
\overline{\psi} \;=\; 2\pi \sum_n n\,|n\rangle\langle n| \; ,\qquad
  {\rm e}^{i 2\pi \overline{Q} /2e}_{}  \;=\; \sum_n |n\rangle\langle n+1| \; , 
\end{equation}
where $\overline{Q} $ is the quasi-charge. Assuming that the junction dynamics is confined to the lowest energy 
band of width $U_0^{}$, we arrive at the Hamiltonian \cite{sz90,ingnaz92}
\begin{equation}\label{josdual}
\begin{array}{rcl}
 H  &=& {\displaystyle  -\, U_0^{}\,\cos\Bigg(\frac{2\pi\overline{Q}}{2e}\Bigg) \,-\, \frac{1}{2\pi}\,2e V_{\rm x}
\, \overline{\psi} } \\[3mm]  
&&  + \; {\displaystyle \sum_\alpha \Bigg[ \frac{\overline{q}_\alpha^{\,2}}{2\overline{C}_\alpha^{}} \,+ \,
\frac{1}{e^2_{}}\frac{1}{2\overline{L}_\alpha^{}}
\left(\frac{\overline{\psi}}{2}
\, -  \, \overline{\psi}_\alpha^{}\right)^2\Bigg] } \; .
\end{array}
\end{equation}
For a current-biased junction we have $V_{\rm x} = R_{\rm Q} I_{\rm x}$.
Since charge and phase are interchanged when turning from the charge representation(\ref{jos2}) to the
phase representation (\ref{josdual}), the environmental effects are now captured 
by the charge autocorrelation function $W_{\overline{\rm Q}}(\tau)= (\pi/e)^2\,\langle [\,\overline{Q}(0) - \overline{Q}(\tau)\,]
\overline{Q}(0)\,\rangle$.
The role of $Z_{\rm t}(\omega)/R_{\rm Q}$ in (\ref{gpsi}) is now taken by $R_{\rm Q} Y(\omega)$, where $Y(\omega)$ is the
admittance. Thus we obtain for a strictly resistive environment 
\begin{equation}\label{wbarq}
W_{\overline{Q}}(\tau) = \frac{2}{\rho} \ln\left[\, \frac{\omega_{\rm R}^{}}{\pi T}\,\sin(\pi T \tau) \,\right]  \; ,
\end{equation}
where $\omega_{\rm R}^{}$ is the required cutoff.

\subsection{Quantum impurity in a Luttinger liquid}

The ground state of 1D interacting spinless electrons is a Tomonaga-Luttinger liquid (TLL), which is characterized by a gapless
collective sound mode. The low-energy modes of the 1D interacting electron liquid is conveniently treated in the framework of bosonization. An extremely useful field-theoretical formulation of a TLL has been given by Haldane \cite{haldane}. 
The creation operator for spinless fermions may be expressed in terms of
bosonic phase field operators $\phi(x,t)$ and $\theta(x,t)$ 
as $\psi^\dagger_{}(x)\propto \sum_{n\,{\rm odd}}^{}\exp\{{\rm i}\,n[k_{\rm F}^{}x+\sqrt{\pi}\,\theta(x)]\}
\exp[{\rm i}\,\sqrt{\pi}\phi(x)]$. At long wave lengths only the terms $n=\pm 1$ are important. They represent the right-
and left-moving parts of the electron field. The bosonic field operators obey the equal-time commutation relation
$[\,\phi(x,t),\,\theta(x',t)\,] = - \,{\rm i}\, {\rm sgn}(x-x')/2$. 

The generic interaction of the TLL universality class is effectively given by a $\delta$-potential. This yields the  
interaction term $H_{\rm I}^{} =U \int {\rm d} x\,{\rm d} y\,\rho(x)\,\delta(x-y)\,\rho(y)$. 
In the TLL, the electron interaction is characterized by a single dimensionless parameter 
$g= 1/\sqrt{1+ U/\pi \upsilon_{\rm F}^{}}$, where $\upsilon_{\rm F}^{}$ is the Fermi velocity. 
Excitations of the liquid are described by the generic harmonic Hamiltonian
\begin{equation}
	H_{\rm L}^{}(g) \;=\; \frac{\upsilon}{2 }\int {\rm d} x\,\left[\,(\partial_x\theta)^2/g\,+\,
g\,(\partial_x\phi)^2\, \right]\; ,
\end{equation}
where $\upsilon =\upsilon_{\rm F}^{}/g$ is the  sound velocity.

Consider now a single strong point-like impurity, or equivalently a weak link or tunnel junction. A hopping or tunneling term which
transfers electrons across the impurity or weak link may be included in the original electronic Hamiltonian by adding
$ H_{\rm I}' = - \Delta [\,\psi^\dagger_{}(x=0^+)\,\psi(x=0^-) + {\rm h.\,c.}\,] $.  Expressed in terms of the bosonic fields,
the hopping term induces a jump of the $\phi$-field at the impurity,
$\overline{\phi} =  \frac{1}{2}[\,\phi(x=0^+) -\phi(x=0^-) \,]$. In the $\phi$-representation, the impurity is then
described by 
\begin{equation}\label{tunint}
H'_{\rm I}(\overline{\phi}) \; =\; -\,\Delta \cos[\,2\sqrt{\pi}\,\overline{\phi} \,+\,
eV_{\rm a}^{} t  \,] \; .
\end{equation}
Here we have included a voltage term. The model is rounded off with addition of the harmonic liquid in the  
right/left ($\pm$) lead. We then arrive at the weak-link Hamiltonian ($\phi$-model) \cite{kf92}
\begin{equation}\label{hphi}
 H_\phi^{} \;=\; H_{\rm L,+}^{}(g)\, +\, H_{\rm L,-}^{}(g)\, +\, H'_{\rm I}(\overline{\phi}) \; . 
\end{equation}

Since the nonlinear degree of freedom is only at $x=0$, we may integrate out the fluctuations of $\phi(x)$ away from $x=0$.
With the Fourier ansatz $ \phi(x,\tau)= T\sum_n \phi(x,\nu_n^{})\,{\rm e}^{-\,{\rm i}\,\nu_n^{} \tau }_{}$,  where $\nu_n^{} = 2\pi T n$
is a Matsubara frequency,
the Euclidean action  is minimized  when $\phi(x,\nu_n^{}) = \overline{\phi}(\nu_n^{})\,\exp(-\,|\nu_n^{} x|/\upsilon)$.   
The resulting Euclidean influence action,
which includes all effects of the excitations in the leads on the tunneling degree of freedom, is
$S^{\rm(E)}_{\rm infl} (\overline{\phi}) = g T\sum_n|\nu_n^{}|\,|\overline{\phi}(\nu_n^{})|^2$.
Average with the weight function ${\rm e}^{-\,S^{\rm(E)}_{\rm infl} (\overline{\phi})}_{}$ yields for the charge correlator
\begin{equation}
\left\langle \mathcal{T}_\tau \,{\rm e}^{ {\rm i}\, 2\sqrt{\pi}\,\overline{\phi}(\tau) }_{} \,
{\rm e}^{ -\,{\rm i}\,2\sqrt{\pi}\,\overline{\phi}(0) }_{} \right\rangle \;=\;
{\rm e}^{ -\,W_{\overline{\phi}}(\tau)}_{}
\end{equation}
with the charge interaction
\begin{equation}
\begin{array}{rcl}
W_{\overline{\phi}}(\tau) &=& {\displaystyle  \frac{2}{g}\, \pi T \sum_n \,\frac{1}{|\nu_n|}
\left( 1-\,{\rm e}^{{\rm i}\,\nu_n^{}t }_{} \right) }\\[4mm]
&=& {\displaystyle  \frac{2}{g} \, \ln\left[\, \frac{\omega_{\rm c}^{}}{\pi T}\,\sin(\pi T \tau) \,\right] }  \; .
\end{array}
\end{equation}

A single weak impurity is modeled by the Hamiltonian
\begin{equation}
H_{\rm sc}^{} = \int {\rm d}x\,U(x)\, \psi^\dagger(x)\psi(x)\; ,
\end{equation} 
where $U(x)$ is the scattering potential. For a point-like
scatterer at $x=0$, the $2k_{\rm F}^{}$-backscattering contribution takes the form  
$H_{\rm sc}^{}(\overline{\theta})= - U_0^{}\cos[\,2\sqrt{\pi}\,\overline{\theta}\,]$, where $\overline{\theta} = \theta(0)$.
With the lead and voltage term added, the weak-impurity or strong-tunneling  Hamiltonian ($\theta$-model) is 
\begin{equation}\label{htheta}
 H_\theta \;=\;  - U_0^{}\cos[\,2\sqrt{\pi}\, \overline{\theta}\,] \,+\, e V_{}^{}\overline{\theta}/\sqrt{\pi} \,+\, 
H_{\rm L}^{}(g) \;.
\end{equation}
One may again integrate out the Luttinger modes away from the scatterer. This leads to the influence action for the scattering mode
$S^{\rm(E)}_{\rm infl} (\overline{\theta}) = (1/g)\, T\sum_n|\nu_n^{}|\,|\overline{\theta}(\nu_n^{})|^2$, and the charge interaction
in the Coulomb gas representation takes the form
\begin{equation}
W_{\overline{\theta}}(\tau) \,=\;  2 g  \, \ln\left[\, \frac{\omega_{\rm c}^{}}{\pi T}\,\sin(\pi T \tau) \,\right]   \; .
\end{equation}

The $\phi$-model is self-dual to the $\theta$-model, and vice versa. In the mapping the regimes of a weak and a strong barrier are 
interchanged. 
We also see that the $\overline{\phi}$- and $\overline{\theta}$-representations 
of the impurity model directly correspond to the above TB- and WB-representations of the Brownian particle in the washboard potential. 
The Kondo parameter $K$ is related to the electronic interaction parameter $g$  by
\begin{equation}
 g \;=\; 1/K \; .
\end{equation} 
Since the correspondences are valid for the Hamiltonians and the effective actions \cite{qds08}, they hold not 
only for transport quantities like the current but also for all quantum statistical fluctuations. 

In addtion, the correspondences also hold for the full counting statistics
of tunneling of edge currents in fractional quantum Hall (FQH) systems, where the fractional filling factor $\nu$
takes the role of $g$ (see e.g. Ref.~\cite{wen90}).
\subsection{One-channel coherent conductor in a resistive electromagnetic environment}
\label{scohcond}
The 1D fermion-boson correspondence suggests that a coherent one-channel conductor under influence of  a
resistive electromagnetic environment can be mapped on a quantum impurity embedded in a TLL. The mapping has been 
analyzed by Safi and Saleur \cite{SafiSaleur}.

A mesoscopic conductor in contact with an electromagnetic environment forms a quantum system violating Ohm's law. 
Transfer of energy from the electrons to the environment leads to dynamical Coulomb-blockade, which reduces the current 
at low voltage.The Luttinger parameter for a coherent conductor in the absence of the electromagnetic environment is $g=1$.  

Following the treatment given in the preceding subsection, and augmenting the phase in the tunneling term (\ref{tunint})
with the phase $\varphi$ of the electromagnetic environment, the relevant weak-tunneling Hamiltonian is found to read
\begin{equation}
\begin{array}{rcl}
H_{\rm wt}^{}  &=& -\, \Delta\cos[\,2\sqrt{\pi}\,\overline{\phi} \,-\, \varphi\,+\, e V_{\rm a}^{}t\,] \\[2mm]  
&&\!\!\!\!\! +\; H_{\rm L,+}^{}(g=1)\, +\, H_{\rm L,-}^{}(g=1)\, +\, H_{\rm env}^{}[{\cal Q},\varphi] \; .
\end{array}
\end{equation}
It is convenient to combine the electronic phase with the electromagnetic phase into the auxiliary phase 
$\overline{\chi}= 2\sqrt{\pi}\, \overline{\phi} -\varphi$. Following the route sketched in the previous subsection, one finds for the 
phase correlation function for a resistive environment
$\langle\, [\overline{\chi}(0)-\overline{\chi}(\tau)]\,\overline{\chi}(0)\,\rangle = W_{\overline{\chi}}(\tau)$ with
\begin{equation}\label{wchi}
W_{\overline{\chi}}(\tau) \,=\;  2 (1+ \rho) \, \ln\left[\, \frac{\omega_{\rm c}^{}}{\pi T}\,\sin(\pi T \tau) \,\right] 
\,+\, 2\rho\,\zeta  \; .
\end{equation}
Here $\rho=R/R_{\rm K}$, and $R_{\rm K} =2\pi/e^2$ is the resistance quantum.

The Hamiltonian of the corresponding strong-tunneling conductor is
\begin{equation}
\begin{array}{rcl}
H_{\rm st}^{} &=&  - \; U_0^{}\cos[\,2\sqrt{\pi}\,\overline{\theta} \,] \,+\, H_{\rm L}^{}(g=1)  \\[2mm]
 && \!\!\!\!\! -\,  
(eV_{\rm a}^{} \,+\,\dot\varphi)\,\overline{\theta}/\sqrt{\pi} \,+\,H_{\rm env}^{}[{\cal Q},\varphi] \; ,
\end{array}
\end{equation}
with $V_{\rm a}$ the applied voltage and $\varphi$ the fluctuating phase of the environment. Upon integrating out the electronic 
modes away from $x=0$, and also the environmental modes for a strictly resistive environment, the effective action takes the form 
$S_{\rm infl}^{\rm (E)}(\overline{\theta})=
(1+\rho)T\sum_n|\nu_n^{}|\,|\overline{\theta}(\nu_n^{})|^2$. Evidently, if we turn to the Coulomb gas representation, this leads to the charge interaction
\begin{equation}\label{wvartheta}
W_{\overline{\theta}}(\tau) \,=\;  \frac{2}{1+\rho} \, \ln\left[\, \frac{\omega_{\rm c}^{}}{\pi T}\,\sin(\pi T \tau) \,\right] 
\,+\, 2 \zeta/\rho \; .
\end{equation}

Again, we have added both in (\ref{wchi}) and in (\ref{wvartheta}) a term which captures the details of the resistive environment
at high frequencies as given in Eq.~(\ref{hifr}). This term
leads to an adiabatic renormalization of the couplings $\Delta$
and $U_0$, as in Eq. (\ref{dressed}).

Thus we find that the two representations of the coherent conductor are self-dual to each other. Furthermore, they can be 
mapped on the models described before. In the mapping to the Brownian particle model and to the quantum impurity 
system we have
\begin{equation}\label{maprel1}
 K  \;\; \hat =\;\; 1/g \;\; \hat = \;\;  1\,+ \, \alpha   \; . 
\end{equation}
These correspondence relations hold both for the weak- and strong-tunneling representations.

\section{Coulomb gas representation}
\label{scgr}

In the sequel, we choose from the above diverse models the tight-binding or weak-tunneling model (\ref{htb}). 
With the duality mapping (\ref{dual1}) and (\ref{dual2}) and the above correspondences between the various models,
subsequent findings can be transferred easily to the  weak-binding regime and to both regimes of the other models.

The change of integration variables, $u_i = 2 \pi T\,\tau_i^{} $, maps the Coulomb gas (\ref{coul1}) with (\ref{corr1}) on the unit circle.
The perturbative series of the partition function (\ref{coul1}) may be written as a power series in the effective fugacity
 \begin{equation}
x = \frac{\Delta}{2 T} \left(\frac{2 \pi \,T}{\omega_c}\right)^K \; .
\end{equation}
The tight-binding series takes the form 
\begin{equation}\label{zexpan}
\mathcal{Z}_{\rm TB}^{}(x,p) \,=\,  1 + \sum_{n\,=\,1}^\infty x^{2n} \mathcal{I}_{2n}(p)\; ,
\end{equation}
where
\begin{equation}\label{icoef1}
\begin{array}{rcl}
\mathcal{I}_{2n}(p) &=& {\displaystyle \frac{2^{-\,2 K n}_{}}{(n!)^2} 
\int_0^{\,2\pi}  \prod_{i\,=\,1}^n \left(\frac{{\rm d}u_i^{}}{\,2\pi} \,\frac{{\rm d}u'_i}{2\pi} \right) }\\[4mm]
&& \!\!\!\! \times \;\;{\displaystyle \left| \frac{\prod_{i<j} \sin ( \frac{u_i-u_j}{2} ) 
\sin (\frac{u'_i-u'_j}{2} ) } {\prod_{i,j}\sin(\frac{u_i-u'_j}{2}) }  
\right|^{2K}\!\! {\rm e}^{{\rm i} p\sum_{i}(u_i -u'_i)}   } \; , 
\end{array}
\end{equation}
and where $p$ represents the scaled physical bias,
\begin{equation} \label{pdef}
p \;=\; {\rm i}\, q  \qquad\mbox{with} \qquad q \;=\; \frac{ \epsilon}{2 \pi\, T}  \;.
\end{equation}
Changing integration variables $z_i = {\rm e}^{{\rm i}\,u_i}_{}$ and $z'_i = {\rm e}^{{\rm i}\,u'_i}_{}$ we get
\begin{equation}\label{ico}
\begin{array}{rcl}
\mathcal{I}_{2n}(p) &=& {\displaystyle      \frac{1}{(n!)^2} \oint \prod_i\left( \frac{ {\rm d}z_i^{}}{2{\rm i}\pi z_i^{}}  
\frac{ {\rm d}z'_i}{2{\rm i}\pi z'_i}  \right)        } \\[5mm]
&\times& {\displaystyle     \frac{[\,\Delta(z)\overline{\Delta(z)}\,]^K  [\,\Delta(z')\overline{\Delta(z')}\,]^K }{
\prod_{i,k}[ \,(1-z_i^{}\bar{z}'_k ) (1-z'_k\bar{z}_i^{} )\,]^K } \,\left( \frac{z_1\cdots z_n}{z'_1\cdots z'_n}\right)^p  }\; ,
\end{array}
\end{equation}
where $\Delta(z) = \prod_{i<k} (z_i^{}-z_k^{})$ is the $n$-variable Vandermonde determinant.

Unfortunately, the multiple integrals in (\ref{icoef1}) or (\ref{ico}) can not be evaluated for general $K$ in the regime
$0<K<\frac{1}{2}$ and for general complex $p$. To advance, we must put $K$ rational and $p$ integer. Integer $p$ may be regarded
as a winding number due to a magnetic charge located at the origin. 
For rational $K$, we may expand the integrand in terms of Jack polynomials \cite{fls95jsp, stanley}
\begin{equation}
\prod_{i,j} \frac{1}{(1-r_i s_j)^K_{}} \;=\;\sum_\lambda b_\lambda(K) P_\lambda(r,K)\,P_\lambda(s,K) \; .
\end{equation}
Here, the function $P_\lambda(r,K)$ is a symmetric polynomial in the set of variables $(r_1,\, r_2,\cdots,\,r_n)$, and
$\lambda = (\lambda_1,\,\lambda_2,\cdots,\,\lambda_n)$ is a partition of an integer, $\lambda_1\le\lambda_2\le\cdots\le\lambda_n$.
For positive integer $p$, we have
\begin{equation}
(z_1\cdots z_n)^p \,P_\lambda(z,K) \;=\; P_{\lambda+p}(z,K) \; , 
\end{equation}
where $\lambda+p$ means the partition $\lambda$ where $p$ columns of length $n$ have been added in the respective Young tableau.  
The multiple integrals in (\ref{ico}) can be executed by use of the orthogonality relation of the Jack polynomials, 
since it involves the Vandermonde determinant. In the end, one  arrives at the totally ordered $n$-fold series expression
\begin{equation} \label{icoef2}
\mathcal{I}_{2n}(p) = \sum_{m_n\,=\,0}^\infty \;\sum_{m_{n-1}\,=\,0}^{m_n}\cdots\sum_{m_1\,=\,0}^{m_2} \; \prod_{j\,=\,1}^n e_j^{}(m_j)
\end{equation}
with
\begin{equation}\label{ejot}
\begin{array}{rcl}
 e_{\!j\,} (m)  &=& {\displaystyle \frac{1}{\Gamma^2(K)}\; 
\frac{\Gamma (j K + m )}{\Gamma ( 1-K +j K+m ) } } \\[3mm]
&& \times\; {\displaystyle \frac{\Gamma (j K+p+m )}{ \Gamma (1-K + j K +p +m ) }  }  \;.
\end{array}
\end{equation}

It has been suggested in Ref.~\cite{fls96jsp} to use (\ref{zexpan}) with (\ref{icoef2}) and (\ref{ejot}) to define 
$\mathcal{Z}_{\rm TB}^{}(x,p)$ for complex $p$. Furthermore, it has been conjectured that this is the unique analytic continuation. Below, we confirm this conjecture in various limiting cases where
analytic expressions are also available by different methods. For instance,  we find from the significant limit
$|p|\to \infty$ that the free energgy $F(x,p) = - T \ln \mathcal{Z}(x,p)$ has the proper behavior
in all orders of $\Delta$, as $T\to 0$.

Consider now the cumulant expansion
\begin{equation}\label{cumexpan}
\ln\mathcal{Z}_{\rm TB}(x,p) \;=\;  \sum_{n\,=\,1}^\infty x^{2n}_{}\, \mathcal{C}_{2n}(p) \; . 
\end{equation}
The first three cumulant coefficients are
\begin{eqnarray}\nonumber
\mathcal{C}_2(p) &=& \mathcal{I}_2(p)  \;,  \\   \label{cumrel}
\mathcal{C}_4(p) &=& \mathcal{I}_4(p) - {\textstyle \frac{1}{2}} \mathcal{I}_2^2(p) \;, \\  \nonumber
\mathcal{C}_6(p) &=& \mathcal{I}_6(p) - \mathcal{I}_2(p)\,\mathcal{I}_4(p) + {\textstyle \frac{1}{3}} \mathcal{I}_2^3(p) \; .
\end{eqnarray}
For the analysis of the limit $|p|\to\infty$ in the cumulant expressions, it will be pivotal to rearrange the multiple sums in 
$\mathcal{C}_{2n}(p)$ into totally ordered sums as in (\ref{icoef2}).
This leads to the split-up
\begin{equation}\label{cum2}
\mathcal{C}_{2n}(p) \;=\; \sum_{m\,=\,1}^n \mathcal{C}_{2n}^{(m)}(p) \; ,
\end{equation}
where $\mathcal{C}_{2n}^{(m)}(p)$ is an $m$-fold ordered sum. The first cumulant is
\begin{equation}\label{c2}
\mathcal{C}_2(p) \;=\; \sum_{j\,=\,1}^\infty e_1^{}(j) \;:
\end{equation}
The second cumulant has two contributions,
\begin{equation}\label{c412}
\begin{array}{rcl}
\mathcal{C}_4^{(1)}(p) &=& {\displaystyle \frac{1}{2}\; \sum_{j\,=\,0}^\infty   e_1^2 (j)\;,   }  \\[3mm] 
\mathcal{C}_4^{(2)}(p) &=&  {\displaystyle \sum_{j\,=\,0}^\infty \sum_{k\,=\,0}^j 
\big[\, e_2^{} (j) - e_1^{} (j)\,\Big]  e_1^{} (k)  \;. }
\end{array}
\end{equation}
The three contributions of the third cumulant are
\begin{equation}\label{c6123}
\begin{array}{rcl}
\mathcal{C}_6^{(1)}(p) &=& {\displaystyle \frac{1}{3} \; \sum_{j\,=\,0}^\infty   e_1^3 (j)\;, } \\[3mm]  
\mathcal{C}_6^{(2)}(p) &=& {\displaystyle \sum_{j\,\ge \,k\,=\,0}^\infty  
\big[\, e_2^{} (j) -e_1^{} (j)\,\big]\,\big[\, e_1^{} (j)+ e_1^{}(k)\,\big]\, e_1^{} (k)  \;,  } \\ 
\mathcal{C}_6^{(3)}(p) &=& {\displaystyle  \sum_{j\,\ge\, k\,\ge\, \ell\,=\,0}^\infty  \Big\{\,  
\big[\, e_3^{} (j)- e_1^{} (j)\,\big]\,  e_2^{} (k)\, e_1^{} (\ell)  }\\     
&&\hspace{3em} {\displaystyle   -\; 2\big[\, e_2^{} (j)- e_1^{} (j)\,\big] \, e_1^{} (k)\,e_1^{} (\ell)\,  \Big\} \;.  }
\end{array}
\end{equation}

The integrals in the expression (\ref{icoef1}), which defines the perturbative coefficients $\mathcal{I}_{2n}(p)$, diverge at short distances when $K\ge 1/2$.
Correspondingly, the multiple series (\ref{icoef2}) for the coefficient $\mathcal{I}_{2n}(p)$ 
diverges in the regime  $K \ge 1/2$ for all $n$.
Interestingly, the coefficient $\mathcal{C}_{2n}(p)$ in the cumulant series (\ref{cumexpan}) is regular  for $K$-values even when 
$\mathcal{I}_{2n}(p)$ is singular.
This is because some of the divergences in the Coulomb integrals are cancelled when taking the connected part 
$\mathcal{C}_{2n}(p)$. The analysis of the ordered $m$-fold sums $C_{2n}^{(m)}(p)$ in (\ref{cum2}) shows that the 
cumulant coefficient $\mathcal{C}_{2n}(p)$ is nonsingular in the range  $0< K < 1-\frac{1}{2n}$ (see below for details).

\section{Nonlinear conductance}
\label{scond}

In Ref.~\cite{fls96jsp} a conjecture was made which relates the nonlinear conductance directly to the partition function $\mathcal{Z}(x,p)$ or to the free energy 
$\mathcal{F}(x,p)$. The conjecture for the
conductance of the TB model (\ref{htb}) is
\begin{equation}\label{cond1}
\begin{array}{rcl}
\mathcal{G}_{\rm TB}\left(x,\epsilon/T\right)\! &=& \!{\displaystyle \frac{K}{2p}\, x\frac{\rm d}{{\rm d}x}\,\ln\left( 
\frac{\mathcal{Z}_{\rm TB}(x,- p)}{\mathcal{Z}_{\rm TB}(x,p)}\right)  } \\[3mm]
&=& \!{\displaystyle \frac{K}{2p } \frac{x}{T}\frac{\rm d}{{\rm d}x}\Big( \mathcal{F}_{\rm TB}(x,p) - \mathcal{F}_{\rm TB}(x,-p)\Big) \;, }
\end{array}
\end{equation}
where $p$ is given in Eq.~(\ref{pdef}). 
The conjecture is based on the fact that the free energy is real only for $p$ integer, and that the continuation of
$C_{2n}(p)$ is not even in $p$, so that the free energy  aquires an imaginary part when $p$ is complex.

The respective relation for the WB model (\ref{hwb}) is
\begin{equation}\label{condwb}
\mathcal{G}_{\rm WB}(x,\epsilon/T) \;=\;  1 \, - \,   \frac{1}{2 K \tilde p}\, x\frac{\rm d}{{\rm d}x}\,\ln\left( 
\frac{\mathcal{Z}_{\rm WB}(x,- \tilde p)}{\mathcal{Z}_{\rm WB}(x,\tilde p)}\right)   \;, 
\end{equation}
where $\tilde p\,=\, p/K  \,=\,{\rm i}\, \epsilon/( 2\pi K T)$.

Use of the perturbative cumulant expansion (\ref{cumexpan}) in the relation (\ref{cond1}) yields the weak-tunneling 
series of the conductance,
\begin{equation}
\mathcal{G}_{\rm TB}(x,\epsilon/T) \;=\; \sum_{n\,=\,1}^\infty \mathcal{G}_{n}(x,\epsilon/T) \; .
\end{equation}
% Here the first term is the normalized maximum conductance in the absence of the impurity. The second term is the perturbative
% backscattering conductance. This term diminishes the  total conductance because of the possibility that the charge is reflected
% at the impurity. 
The perturbative tunneling contribution of order $n$ is
\begin{equation}\label{bsc1}
\mathcal{G}_{n}(x,V/T) \;=\;  \frac{K}{{\rm i}\,q}\,n\big[\, \mathcal{C}_{2n}(-\,{\rm i}\,q) \,-\,\mathcal{C}_{2n}({\rm i}\,q)
\, \big] \,x^{2n}_{}\; , 
\end{equation}
where $q= \epsilon/(2\pi T)$. We now prove the conjecture in diverse parameter regimes.

\subsection{Strong barrier limit}

The series that defines $\mathcal{C}_2(p)$ can be summed in analytic form
\begin{equation}
\begin{array}{rcl}
\mathcal{C}_2(p) &=& {\displaystyle \frac{1}{\Gamma^2(K)}\,\sum_{j\,=\,0}^\infty 
\frac{\Gamma(K+j)\,\Gamma(K+p+j)}{\Gamma(1 +j)\,\Gamma(1+p+j)} }\\[5mm]
&=& {\displaystyle \frac{\sin(K \pi )\,\Gamma(1-2 K)}{\pi} \;\frac{\Gamma(K+p)}{\Gamma(1-K+p)} } \; .
\end{array}
\end{equation}
With this form the perturbative conductance (\ref{bsc1}) in order $\Delta^2$ is found as
\begin{equation}\label{bsc2}
\mathcal{G}_1 \;=\; K \,\left(\frac{\Delta}{2 T}\right)^2 \left(\frac{2\pi T}{\omega_{\rm c}^{}}\right)^{2K}\,
\frac{\sinh(\frac{\epsilon}{2T})}{\frac{\epsilon}{2T}}\,\frac{|\Gamma( K + {\rm i}\,\frac{\epsilon}{2\pi T} )|^2}{\Gamma(2K)} \;.
\end{equation}

Correspondingly, the perturbative conductance through a voltage-biased Josephson junction arising from incoherent Cooper pair tunneling is
\begin{equation}\label{pertcond}
\mathcal{G}_1 \;=\; \rho\,\left(\frac{E'_{\rm J}}{2 T}\right)^2 
\left(\frac{2\pi^2\rho  T}{E_{\rm c}^{}}\right)^{2\rho}\,
\frac{\sinh(\frac{ e V}{T})}{\frac{e V}{T}}\,\frac{|\Gamma( \rho + {\rm i}\,\frac{e V}{\pi T} )|^2}{\Gamma(2\rho)} \;.
\end{equation}
The expressions (\ref{bsc2}) and (\ref{pertcond}) are exactly the same as the ones obtained from the non-equilibrium real-time approach 
\cite{qds08,ing94}.

\begin{figure}[ht!]
\vspace{0cm}
\begin{center}
\includegraphics[scale=1.0]{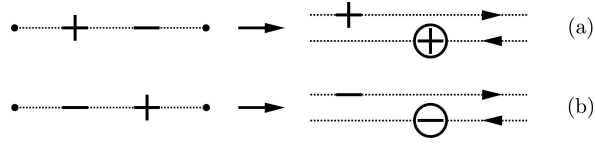}
\end{center}
\vspace{-1.3em}

\caption{\label{f1} Imaginary-time charge diagrams representing $\mathcal{Z}_1(\mp p)$ or the connected part 
$\mathcal{F}_1(\mp p)$ (left), and analytic continuation to real-time forward/backward path (right). Diagram
(a) represents the forward rate $k_{1,1}^{(+)}$, and diagram (b) the backward rate $k_{1,1}^{(-)}$. 
Each graph has fixed charge order along the forward/backward path. Enqueuing the double path (a) into a single path yields two
contributions, $\; + \;\oplus\; $ and  $\;\oplus\;+\;$, while diagram (b) yields  $\; - \;\ominus\; $ and  $\;\ominus\;-\;$. 
These are taken into account by the integration limits $-\,{\rm i}\,\infty$ and $\,{\rm i}\,\infty$ in (\ref{imfint}). }
\end{figure}

Let us study the equivalence in some detail.
In the weak tunneling limit, the normalized conductance is given in terms of the golden rule
transition rates $k_{1,1}^{(\pm)}$ representing incoherent forward/backward tunneling transitions to adjacent wells 
(see Section \ref{snoise} for notation).  We have
\begin{equation}\label{condgr}
\mathcal{G}_1 \;=\; {\displaystyle \frac{K}{qT}\,\left( k^{(+)}_{1,1} \,-\,k^{(-)}_{1,1} \right)  }  \; ,
\end{equation}
where $q$ is given in (\ref{pdef}).
The rates may be calculated directly from the standard real-time double-path representation in order $\Delta^2$, which is 
pictorially sketched in the right diagrams of Fig.~\ref{f1}. Alternatively, one may calculate the rates with the standard
${\rm Im}\,\mathcal{F}$-method \cite{qds08}. Here, the free energy of order $\Delta^2$,
$\mathcal{F}_1(\mp p)$, (see left diagrams of Fig.~\ref{f1}) is analytically continued to a complex value. 
The imaginary part of the free energy is obtained by distorting the integration path of the charge correlation integral 
from imaginary to real time. In this way we get
\begin{equation} \label{imfint}
k^{(\pm)}_{1,1}\;=\; -\,2\,
{\rm Im}\,\mathcal{F}_1(\mp p) \; = \;  \frac{x^2T}{\pi\,{\rm i}}\int_{0^+ -\,{\rm i}\,\infty}^{0^+ +\,{\rm i}\,\infty}
{\rm d}\upsilon\,\frac{{\rm e}^{\mp\,{\rm i}\,2p \upsilon}_{} }{(2\sin \upsilon)^{2K}} \; .
\end{equation}
Taking either road we obtain the unique expression \cite{qds08} 
\begin{equation}\label{krel}
k_{1,1}^{(\pm)} \;=\; x^2 T \left[\, {\rm e}^{-\,{\rm i}\,K \pi }_{}\,J_1(K,\mp \,p) \, +\,
{\rm e}^{{\rm i}\, K\pi }_{}\,J_1(K,\pm \,p)  \,\right]   \;  ,
\end{equation}
where $J_1(K,p)$ is the real-time noise correlation integral
\begin{equation} \label{intreal}
\begin{array}{rcl}
J_1(K,p) &=& {\displaystyle \frac{1}{\pi} \!\int_0^\infty \!\!{\rm d}u\,\frac{{\rm e}^{-2 p u}_{}}{(2 \sinh u)^{2K}} } \\[4mm]
&=& {\displaystyle  \frac{\Gamma(1-2K)\,\Gamma(K+p)}{2\pi\,\Gamma(1-K+p)} }\; .
\end{array}
\end{equation}
Upon combining (\ref{condgr}) with (\ref{krel}) we get
\begin{equation}\label{condreal}
\mathcal{G}_1 \;=\; \frac{ 2K}{p} \,x^2 \sin(K \pi ) \,\left[\, J_1(K,-p) \,-\, J_1(K,p) \,\right] \; ,
\end{equation}
which is in agreement with the expression (\ref{bsc2}).

Consider next the conjecture (\ref{cond1}) with (\ref{zexpan}) and (\ref{icoef1}). This yields the expression
\begin{equation}\label{condim}
\mathcal{G}_1 \;=\; \,\frac{ K}{p} \,x^2 \,\left[\, H_1(K,-p) \,-\, H_1(K,p) \,\right] \;.
\end{equation}
The function $H_1(K,p)$ coincides with the cumulant $C_2(p)$. It is instructive that $H_1(K,p)$ can be written as the 
imaginary-time noise correlation integral
\begin{equation}
H_1(K,p)  \;=\; \frac{1}{\pi} \,\int_0^\pi \!\!{\rm d}\upsilon\, \frac{  {\rm e}^{\,2{\rm i}p \upsilon}_{}}{ 
(2 \sin \upsilon)^{2K}}  \; .
\end{equation}
The analytically continued integrand of (\ref{intreal}), ${\rm e}^{-2 p z}_{}/(\sinh z)^{2K}$, 
is free of singularities in the half-strip $ 0<{\rm Re}\,z<\infty\;,\; \;0>{\rm Im}\,z > - \pi\;$ of the com\-plex plane 
$z=u\,-\,{\rm i}\,\upsilon$. Accordingly, the closed contour integral along the edges of the half-strip vanishes.
This entails that the noise integrals $H_1(K,p)$ and $J_1(K,p)$ satisfy the relation
\begin{equation}\label{reimrel1}
H_1(K,p) \;=\; 2\,{\rm e}^{\,{\rm i}p\pi}_{} \,\sin[(p+K)\pi]\,J_1(K,p)  \;.
\end{equation}
For integer $p$, this relation reduces to
\begin{equation}\label{reimrel2}
H_1(K,p) \;=\;  2\sin(K\pi)\,J_1(K,p)\; , \qquad  p \in {\rm integers}  \; .
\end{equation}
Hence the expressions (\ref{condreal}) and (\ref{condim}) coincide for integer $p$. 
Evidently, the result (\ref{condreal})  of the real-time calculation is also valid for complex $p$. This confirms
that the expression (\ref{condim}), which is based on the thermodynamic method and has been calculated for integer $p$, 
can be analytically continued to complex $p$.

Unfortunately, the real-time approach is not practical for perturbative contributions $\mathcal{G}_n(x,\epsilon/T)$ with $n\ge 2$,
since the relevant Coulomb gas multiple integrals for charges  distributed on the forward/backward paths can not be 
carried out in analytic form. The loophole now is to take the imaginary time-route to calculate the free energy for integer $p$, as sketched in 
Section \ref{scgr}. With the conjecture that the resulting expression (\ref{zexpan}) with (\ref{icoef2}) 
can be analytically continued to complex $p$, we then get  the conductance (\ref{cond1}). Let us now see whether this line of
argument is correct in all orders of $x$.

\subsection{The case $K\ll 1$}

In the regime $K\ll 1$, the leading contribution to the coefficient $\mathcal{I}_{2n}(p)$ in the multiple series (\ref{icoef2}) is the summand labelled with $m_1=m_2=\cdots=m_n=0$. 
Furthermore, for $K\ll 1$ the product term $e_j(0)$ given in (\ref{ejot}) reduces to the form $e_j(0) = K/[\,j\,(K+p)\,]$. With these truncations the series coefficient $ \mathcal{I}_{2n}(p)$ takes the concise form
\begin{equation}
\mathcal{I}_{2n}(p) \;=\; \prod_{j\,=\,1}^n\,e_j(0) \:=\; \frac{1}{n!}\,\frac{\Gamma(1+p/K)}{\Gamma(1+n+p/K)} \; .
\end{equation}
Interestingly, the resulting perturbative series of the partition function (\ref{zexpan}) can be summed in closed form, 
\begin{equation}
\begin{array}{rcl}
 \mathcal{Z}_{\rm TB}(x,p)\!\! \!&=& \!\!\!{\displaystyle  
\Gamma(1+p/K)\,x^{-p/K}\sum_{n\,=\,0}^\infty \frac{x^{2n+p/K}}{n!\Gamma(1+n+p/K)} } \\[5mm]
 \! \! &=& \! \!\Gamma(1+p/K)\, x^{-p/K} I_{p/K}(2x) \;.
\end{array}
\end{equation}
The function $I_{\nu}(z)$ is a modified Bessel function with index $\nu$. Applying functional relations of the Bessel functions \cite{abr} the tunneling 
susceptibility $ {\rm d}\mathcal{F}/{\rm d}x$ takes the analytic form
\begin{equation}
\frac{{\rm d}}{{\rm d}x}\mathcal{F}_{\rm TB}(x,p) \;=\; -\,T  \frac{I_{1 +p /K}(2x)}{I_{p/K}(2x)} \; .
\end{equation}
With this the expression (\ref{cond1}) for the nonlinear conductance takes the concise analytic form
\begin{equation}\label{cond2}
\mathcal{G}(x,\epsilon/T) \,=\, \frac{K}{q}\, 2 x\, {\rm Im}\, \frac{I_{1-{\rm i}\,q/K}(2x)}{I_{-{\rm i}\,q/K}(2x)}\;,\qquad
q =\frac{\epsilon}{2\pi T} \; .
\end{equation}
Upon employing recursion relations of the Bessel functions, this expression can be transformed into
\begin{equation}\label{cond3}
\mathcal{G}(x,\epsilon/T) \;=\; 1 \,-\, \frac{\sinh(\pi q/K)}{\pi q/K} \frac{1}{|I_{\,{\rm i}\,q/K}(2x)|^2 } \; .
\end{equation}
This form reduces to the linear conductance
\begin{equation}\label{cond4}
\mathcal{G}_{\rm lin}(x) \,=\, \mathcal{G}(x,0) \,=\, 1\,-\, \frac{1}{I_0^{\,2}(2x)} \; .
\end{equation}

With the substitution $2x\to E_{\rm J}/T$, $\epsilon\to 2 eV$, and $K\to\rho$, the expression
(\ref{cond2}) corresponds to the result of Ivanchenko and Zil'berman for the conductance of a classic overdamped Josephson junction
\cite{iva}. For the standard derivation of the expression (\ref{cond2}) within  a real-time  Coulomb gas approach, resulting in an infinite
continued fraction expression, we refer to Refs.~\cite{gip98,zwerger87,qds08}.

In the weak-tunneling model (\ref{htb}), the regime $K\ll 1$  corresponds to the classical regime of the 
Brownian particle model (\ref{hwb}), 
as  the damping parameter $K_{\rm WB} =1/K$ is very large. The respective conductance is independent of $K_{\rm WB}^{}$ and is
\begin{equation}\label{cond5}
\begin{array}{rcl}
\mathcal{G}(x,\epsilon/T) &=&{\displaystyle 1 \,-\, \frac{2x}{q}\,{\rm Im}\, \frac{I_{1-{\rm i}\,q}(2x)}{I_{-{\rm i}\,q }(2x)} }\\[4mm]
&=& {\displaystyle   \frac{\sinh(\frac{\epsilon}{2T})}{\frac{\epsilon}{2T}}\,\frac{1}{|I_{{\rm i}\,q}(2x)|^2} } \;, 
\end{array}
\end{equation}
as follows from the selfduality (\ref{dual2}). In the strict classical limit we have $2x =U_0/T$ .

Weak adiabatic quantum fluctuations may be taken into account (i) in the WB model by an effective corrugation strength 
$U_0 \to U^\ast_0= U_0 \,(2\pi T/\omega_{\rm c}^{})^{1/K}_{}$, and (ii) in the model (\ref{jos2}) by an effective Josephson energy 
$E_{\rm J}\to E^\ast_{\rm J}= E'_{\rm J}\, (2\pi T/\omega_{\rm R}^{})^\rho$ \cite{gip98}, respectively.

Consider next the zero temperature limit of the expression (\ref{cond5}). As $T\to 0$, both the argument and the modulus of the
Bessel function  become very large. In this case the uniform asymptotic expansion of the modified Bessel function  applies \cite{abr}. 
The resulting expression for the nonlinear conductance of the weak-corrugation model (\ref{hwb}) is
\begin{equation}\label{cond6}
\mathcal{G}|_{T\,=\,0} \,=\, \Theta(\epsilon - 2\pi U_0) \sqrt{ 1- \left( \frac{2\pi U_0}{\epsilon}\right)^2 } \; .
\end{equation}

The expressions (\ref{cond5}) and (\ref{cond6}) coincide with the solutions of the classical Smoluchowski equation 
in corresponding regimes \cite{risken}. For $\epsilon< 2\pi U_0$, the conductance (\ref{cond6}) is zero because the sliding motion of the overdamped particle comes to rest at locations with  zero slope. The particle rests there forever, since thermal and quantum
fluctuations are absent in the classical regime at $T=0$. For $\epsilon > 2\pi U_0$, the potential in (\ref{hwb}) is sloping down everywhere, so that the particle is continuously sliding.

Transcribing (\ref{cond6}) into the regime of classical phase diffusion in the Josephson junction, which corresponds 
to the regime $\rho<<1$, yields for the nonlinear conductance at $T=0$ 
\begin{equation}
\mathcal{G}|_{T\,=\,0} \,=\, \Theta(V - R I_{\rm c})\,\left[ \,1 -  \sqrt{ 1- \left( R I_{\rm c}/V  \right)^2 } \, \right]  \; ,
\end{equation}
where $I_{\rm c} =2 e E_{\rm J}$ is the critical current.

In conclusion, our analysis shows that the above imaginary-time approach to nonequilibrium transport in the limit $K\to 0$ 
(and $K\to \infty$ in the dual model) reproduces all
known expressions calculated from  an elaborate  real-time method.

\subsection{The limit $T\rightarrow 0$}

The limit $T\to 0$ constitutes a crucial proof whether the analytic continuation from integer $p$ to complex $p$ in the expression (\ref{zexpan}) with (\ref{icoef2}) is correct for general $K$ in all orders of $x$. 

For general $p$, the $m$-fold ordered sum in $\mathcal{C}_{2n}^{(m)}(p)$ can not be done in analytic form. The simplest term
$\mathcal{C}_{2n}^{(1)}(p)$ is a generalized hypergeometric series of the form $_{2n}F_{2n-1}(\cdots;\cdots;1)$. 

As $|p|\to\infty$, the ordered sums in $\mathcal{C}_{2n}^{(m)}(p)$
turn into ordered integrals which can be solved in analytic form. With the substitution $k \to p \,u$, we have the mapping 
\begin{equation}\label{asy1}
\sum_k \cdots \;\to\;  p \int {\rm d}u \,\cdots \;,
\end{equation}
and the asymptotic expansion of the $e_j^{}(pu)$ emerges as
\begin{equation}\label{asy}
\begin{array}{rcl}
 e_j^{}(p u) \!&=& \!{\displaystyle  \frac{p^{2K-2} }{\Gamma^2(K)} \,\big( u\,U\big)^{K-1} \left\{\,1 + \frac{b_1(j)}{p} \,\frac{u+U}{u\,U} \right.}
 \\[4mm] 
&& \!\!\!\!+{\displaystyle\left.   \frac{1}{p^2}\left[b_{21}(j)\frac{u^2+U^2}{u^2\, U^2} + b_{22}(j)\frac{1}{u\,U}\right]   
 +   \cdots \!  \right\}   } \; ,
\end{array}
\end{equation}
where $U=1+u$. The expansion coefficients are
\begin{eqnarray}\nonumber
b_1(j) &=& {\textstyle \frac{1}{2}}\,K(K-1)(2j-1) \;, \\[1mm] \label{expancoef}
b_{21}(j) &=& {\textstyle \frac{1}{24}}\, K(K-1)(K-2)\big[\,3K(2j-1)^2-1 \,\big]  \;,  \\[1mm]    \nonumber
b_{22}(j) &=& {\textstyle \frac{1}{4}}\, K^2(K-1)^2 (2j-1)^2 \; .
\end{eqnarray}

From this we find that each term of $\mathcal{C}_{2n}^{(m)}(p)$ behaves asymptotically as $p^{(2K-2)n+m}$. However, there are formidable cancellations by adding up the individual terms, as we immediately see from the expressions (\ref{c2}) - (\ref{c6123}). As a result, as far as
the cumulant $\mathcal{C}_{2n}(p)$ is concerned, subleading terms in the curly bracket of (\ref{asy}) become the leading ones. 
The analysis shows that there is  an extra factor 
\begin{equation}\label{sublead}
(K/p)^{m-1}_{} \;,
\end{equation}
 which results from the relevant subleading terms. Thus, in reality, the 
asymptotic power law is $\mathcal{C}_{2n}^{(m)}(p) \propto p^{(2K-2)n +1}_{}$, and hence $\mathcal{C}_{2n}(p) \propto p^{(2K-2)n +1}_{}$.

Consider now first the partial cumulant coefficient 
\begin{equation}
\mathcal{C}_{2n}^{(1)}(p) \,\equiv\;
\frac{1}{n}\frac{1}{\Gamma^{2n}(K)} \sum_{j\,=\,0}^\infty \,\left(\frac{\Gamma(K+j)\, \Gamma(K+p+j)}{\Gamma(1+j)\,\Gamma(1+p+j)}
\right)^n  \; .
\end{equation}
In the asymptotic regime $|p|\to\infty$ we have
\begin{eqnarray} \label{cumzero}
C_{2n}^{(1)}(p)
\!\!&=& \!\! \frac{p^{2(K-1)n +1}}{n\,\Gamma^{2n}(K)}\int_0^\infty \!\! {\rm d}u\,[\,u(1+u)\,]^{(K-1)n} \\ \nonumber
\!\!&=&\!\! \frac{p^{2(K-1)n +1}}{n\,\Gamma^{2n}(K)} \frac{\Gamma(1-n+n K)\,\Gamma(2n-1-2n K)}{\Gamma(n- n K)}  \; .
\end{eqnarray}
The integral in the first line of (\ref{cumzero}) is convergent in the regime $1- \frac{1}{n} < K < 1-\frac{1}{2n}$. 
On the other hand, the total cumulant $C_{2n}(p)$ is regular down to $K=0$, as follows from the integrals in (\ref{icoef1}), i.e.,
$0 < K < 1-\frac{1}{2n}$. In comparison with $C_{2n}(p)$, the expression (\ref{cumzero}) has $n-1$ additional singularities located at
$K = \frac{m}{n}$, where $m=1,\,2,\cdots,\,n-1$. The spurious singularities are contained in the term $\Gamma(1-n+nK)$.
The analysis now gives that the  sum of all partial cumulants in (\ref{cum2}) gives $C_{2n}^{(1)}(p)$ 
times a polynomial $a_n(K)$ which is of order $\,n-1$ in $K$,
\begin{equation}\label{cumgen1}
C_{2n} (p) = a_{n}^{}(K) \, C_{2n}^{(1)}(p) \; .
\end{equation}
The polynomial $a_n(K)$ provides zeros just at the locations of the spurious singularities, so that $C_{2n}(p)$ is indeed regular in the range
$0\le K < 1-\frac{1}{2n}$. With the limiting value $a_n(0)=1$,
which follows from (\ref{sublead}), we thus get the unique expression
\begin{equation} \label{acoef}
\begin{array}{rcl}
a_{n}(K) &=& {\displaystyle \frac{(-1)^{n-1}}{(n-1)!}\,\prod_{m\,=\,1}^{n\,-\,1} (n K -m)} \\[4mm]       
 &=& {\displaystyle \frac{(-1)^{n-1}}{(n-1)!}\, \frac{\Gamma(n K)}{\Gamma(1-n + n K)} } \; .
\end{array}
\end{equation}

Insertion of (\ref{acoef}) into (\ref{cumzero}) yields the asymptotic form of the cumulant coefficients in the regime $|p|\to \infty$,
\begin{equation}\label{cumres}
C_{2n}(p) \;=\; \frac{(-1)^{n-1}}{n !}\, \frac{\Gamma(n K)}{\Gamma^{2n}(K)} \frac{\Gamma(2n-1- 2n K)}{\Gamma(n - n K)}
\,p^{2(K-1)n +1} \; .
\end{equation}
In order to corroborate these considerations and the result (\ref{cumres}), we calculate $C_4(p)$ and $C_6(p)$ 
in \ref{sapp}  explicitly.

The cumulant $C_{2n}(p)$ has simple poles at $K=1- \frac{1}{2n}+\frac{m}{n}$, where $m=0,\,1,\,2,\cdots$. These poles are 
due to the Gamma function term $ c_n^{}(K) =  \Gamma(2n-1- 2n K)/\Gamma(n - n K)$ in $C_{2n}(p)$.

Consider next the analytic continuation provided in the expression (\ref{bsc1}). Most interestingly, this produces  
a trigonometric factor $\cos[\,(\pi(1- K)n\,]$, which has zeros just at the poles of $c_n^{}(K)$.
As a result, the perturbative conductance contribution $\mathcal{G}_n(x,\epsilon/T)$ turns out to be regular for all $K>0$. 
It may be written as
\begin{equation}\label{pertcontr}
\mathcal{G}_n \;=\; \frac{(-1)^{n-1}}{n!}\,\frac{\Gamma(\frac{3}{2})\Gamma(1+n K)}{\Gamma^{2n}(K)\Gamma(\frac{3}{2} - n+n K)}
\left( \frac{q}{2}\right)^{2(K-1)n} x^{2n}  \; .
\end{equation} 
A crucial point now is that temperature cancels out in the combined expression $q^{2(K-1)n}_{} x^{2n}_{}$. 

Finally, it is useful to combine the tunneling coupling $\Delta$ and cutoff $\omega_{\rm c}^{}$ to the 
universal Kondo scale $\epsilon_0^{}$ as \cite{qds08}
\begin{equation}\label{epstb}
\epsilon_0^{2-2K} \;=\;  2^{2-2K}_{} \frac{\pi^2}{\Gamma^2(K)}\,\frac{\Delta^2}{\omega_c^{2K}} \; .
\end{equation}
With use of this scale in the expression (\ref{pertcontr}), the weak-tunneling series of the conductance at zero temperature and general $K$ takes the concise form
\begin{equation}\label{tbseries}
\mathcal{G}(\epsilon,\epsilon_0^{},K) \; =\; \sum_{n\,=\,1}^\infty \frac{(-1)^{n-1}}{n!}\,\frac{\Gamma(\frac{3}{2})
\Gamma(1+nK)}{\Gamma(\frac{3}{2}- n + nK )} \left( \frac{\epsilon_0^{}}{\epsilon}\right)^{2(1-K)n}  \; .
\end{equation}
This is exactly the expression for the conductance at $T=0$ found with the thermodynamic Bethe ansatz by Fendley, Ludwig and 
Saleur \cite{fls95prb}. The series (\ref{tbseries}) is appropriate for $K<1$ in the regime $\epsilon>\epsilon_{\rm cr}^{}$,
where $\epsilon_{\rm cr}^{} = \sqrt{|1-K|}K^{K/[2(1-K)]}_{}\,\epsilon_0^{}$. 

For $K<1$ and $\epsilon<\epsilon_{\rm cr}^{}$, the appropriate starting point would be the relation (\ref{condwb}). In this case, the resulting expression for the conductance is the strong-tunneling series
\begin{equation}\label{wbseries}
\mathcal{G}(\epsilon,\epsilon_0^{},K) \; =\; 1 - \sum_{n\,=\,1}^\infty \frac{(-1)^{n-1}}{n!}\,\frac{\Gamma(\frac{3}{2})
\Gamma(1+n/K)}{\Gamma(\frac{3}{2}- n + n/K )} \left( \frac{\epsilon_0^{}}{\epsilon}\right)^{2(1-1/K)n}  \; .
\end{equation}
Expressed in terms of the parameters of the WB model, the Kondo scale $\epsilon_0^{}$ is 
\begin{equation}\label{epswb}
\epsilon_0^{2-2/K} \;=\; (2K)^{2-2/K}_{}\,\frac{\pi^2}{\Gamma^2(1/K)}\,\frac{U_0^2}{\omega_{\rm c}^{2/K}} \; .
\end{equation}
Here the extra factor $K^{2-2/K}$ compared to (\ref{epstb}) accounts for the mapping $\epsilon\to\epsilon/K$ in the 
self-duality relation (\ref{dual2}).

For $K>1$ and fixed $\epsilon$, the regimes of weak tunneling, Eq.~(\ref{tbseries}), and strong tunneling, Eq.~(\ref{wbseries}), 
are exchanged.

We conclude this subsection with the remark that there are contour integral representations of the conductance  of which (\ref{tbseries}) is the Taylor series and (\ref{wbseries}) the asymptotic series,
or vice versa \cite{weiss96,fs98,qds08}.

\section{Rates, current and noise}
\label{snoise}

The conductance $\mathcal{G}$ and the current $\langle I \rangle = \epsilon \mathcal{G}/(2\pi K)$ are first moments of the 
transition rates,
\begin{equation} \label{curr}
\langle I \rangle  \;=\; \sum_{n=1}^\infty n\,\left(k_n^{(+)} - k_n^{(-)}\right)\; .
\end{equation}
Higher moments of the current give information on the statistical fluctuations of the transport process,
\begin{equation}\label{noise}
\langle I^{(m)}\rangle_{\rm c}  \;=\; \sum_{n\,=\,1}^\infty  n^m_{}\,
\left( k_n^{(+)}\, +\, (-1)^m  k_n^{(-)} \right)      \; .
\end{equation}
A knowlegde of all the connected moments (\ref{noise})  allows then a reconstruction of the 
population probability distribution. 

The dynamics of the population probability $P_n(t)$ of site $n$ is governed by the master equation
\[
\dot P_n(t) = \sum_{\ell\,=\,1}^\infty \left[\,  k_\ell^{(+)}P_{n-\ell}(t) +    k_\ell^{(-)}P_{n+\ell}(t)
-\left( k_\ell^{(+)}+k_\ell^{(-)} \right) P_{n}(t) \,\right]  \; .
\]
The characteristic function $\tilde P(\lambda,t) = \sum_n {\rm e}^{{\rm i}\,\lambda n}\,P_n(t)$ with initial state $P_n(0)=\delta_{n,0}^{}$
is found from the master equation to read 
\begin{equation}\label{charfun}
 \tilde P(\lambda,t) \;=\; \prod_{n\,=\,1}^\infty \exp\left[ t({\rm e}^{{\rm i}\,\lambda n}\,-\,1)k_n^{(+)} \,+\,   
 t({\rm e}^{-\,{\rm i}\,\lambda n}\,-\,1)k_n^{(-)}    \right] \; .
\end{equation}
The moments of the probability distribution follow from the characteristic function by differentiation,
\begin{equation}
\langle n^{(m)}_{}(t)\rangle\;=\;\sum_n n^m_{}\,P_n(t) \;=\; \left(-\,{\rm i}\,\frac{\partial}{\partial \lambda}\right)^m\,
\tilde P (\lambda,t)\Big|_{\lambda\,=\,0} \; .
\end{equation}
The cumulant expansion 
\begin{equation}
\frac{\partial}{\partial t}\,\ln \tilde P(\lambda,t)\;=\; \sum_{n\,=\,1}^\infty\frac{({\rm i}\,\lambda)^m}{m!} \langle I^{(m)}\rangle_{\rm c}
\end{equation}
leads us to the connected moments (\ref{noise}) of the current.

It is evident from (\ref{noise}) that knowledge of all moments is tantamount to knowledge of all rates $k_n^{(\pm)}$ 
in all orders of $x$ or $\Delta$.

The rate $k_n^{(\pm)}$ is built up by paths
which start out at an arbitrary initial state of the reduced density matrix (RDM), say $(0,0)$, 
then may run through any number off-diagonal 
and diagonal states of the RDM, and finally end in diagonal state $(\pm n,\pm n)$. 
The minimal number of moves, each of weight $\Delta/2$, is $2n$. 
Since the paths may make arbitrary detours, the transition rate $k_n^{(\pm)}$ is a series of partial rates 
$k_{n,\ell}^{(\pm)}$, where the second index 
indicates that the transition consists of $2\ell$ moves, or $2\ell$ charges in the charge representation,
\begin{equation}\label{rateseries}
 k_n^{(\pm)} \;=\; \sum_{\ell\,=\,n}^\infty k_{n,\ell}^{(\pm)} \;= \; \sum_{\ell\,=\,n}^\infty R_{n,\ell}^{(\pm)}\,x^{2\ell}_{} \; .
\end{equation}
In the last form, the dependence on the effective fugacity $x$ is exposed.
For paths, which have interim visits of diagonal states, the reducible part must be subtracted ( see Ref.~\cite{hbaur}).

\begin{figure}[ht!]
\vspace{0cm}
\begin{center}
  \includegraphics[scale=1.0]{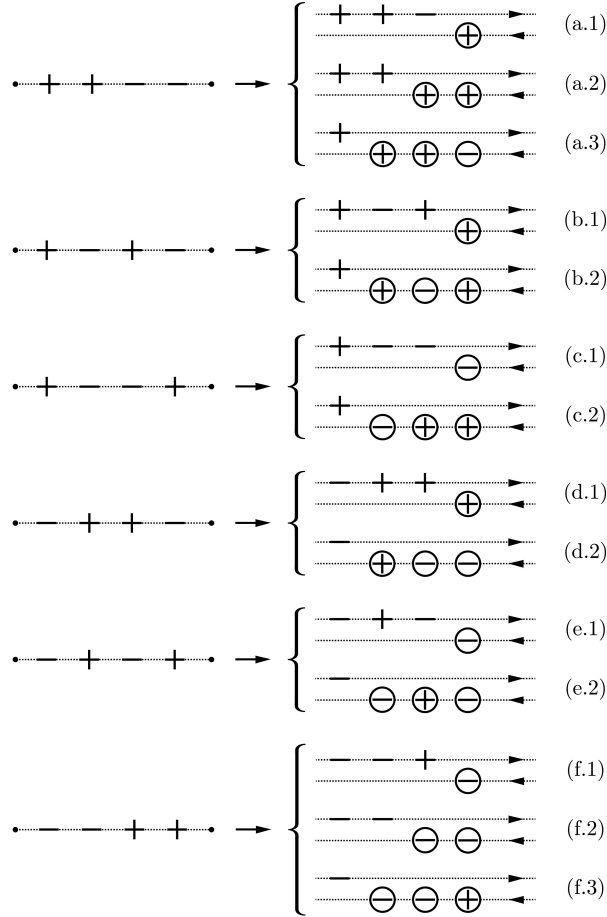}
\end{center}
\caption{ \label{f2} The six imaginary-time charge diagrams representing $\mathcal{Z}_2$ or the connected part 
$\mathcal{F}_2$ (left),  and the various possibilities to partition the charges on the forward/backward path. A charge on the 
backward path changes sign when the time axis is reversed. The encircled charges are the charges $\{ \upsilon_i\}$ in (\ref{infl}). 
Six diagrams each
contribute to $k_1^{(+)}$ and to $k_1^{(-)}$, and one diagram each represents $k_2^{(+)}$ and $k_2^{(-)}$.
The other sixteen charge sequences, which all  contribute to $k_0^{(\pm)}$, are omitted.
Upon enqueuing the various charge sequences along the forward/backward path into a single path, the graphs $(a.2)$ and $(f.2)$ each split up into 6 
different time-ordered charge sequences, and each of the residual 12 graphs splits up into 4 different time-ordered charge sequences.  }
\end{figure}

The partial rates $k_{n,\ell}^{(\pm)}$ may be calculated via analytical continuation of the $\ell$th cumulant of the 
partition function  $\mathcal{Z}$.
In the analytical continuation the $2\ell$ charges representing the cumulant $C_{2\ell}$ may be arbitrarily partitioned on the 
forward and backward path, as sketched for the case $\ell=2$ in Fig.~\ref{f2}.
The charge correlation factor for $m$ time-ordered charges $\{u_j=\pm 1\}$ on the forward path $q(t)$ and 
$\ell$ time-ordered charges $\{ \upsilon_i=\pm 1\}$ on the backward path $q'(t')$ is 
\begin{equation} \label{infl}
\begin{array}{rcl}
\mathcal{F}_{\rm corr}[q_m,q'_\ell] &=&{\displaystyle  \exp\left\{ \; \sum_{j\,=\,2}^m \sum_{i\,=\,1}^{j-1}\, 
u_ju_i Q(t_j-t_i) \right.} \\[5mm]  
      && \qquad\;\;+\;{ \displaystyle  \left. \sum_{j\,=\,2}^\ell\sum_{i\,=\,1}^{j-1}\, \upsilon_j \upsilon_i Q^\ast(t'_j-t'_i)  \right.  }   \\[5mm]  
       &&\qquad\;\;  - \; {\displaystyle \left.   \sum_{i\,=\,1}^\ell \sum_{j\,=\,1}^m \,\upsilon_i u_j Q(t'_i -t_j)  \;  \right\} } \\[6mm]
&&\times\;{\displaystyle  \exp\left\{{\rm i}\,\epsilon\left( \sum_{j\,=\,1}^m u_j\, t_j\,-\, 
\sum_{i\,=\,1}^\ell \upsilon_i\, t'_i\right)  \;   \right\} } \;,
\end{array}
\end{equation}
where $Q(t) = 2 K \ln[(\omega_{\rm c}/\pi T)\sinh(\pi T t)]+{\rm i}\,\pi K\,{\rm sgn}(t)$, and where the last line represents the
correlations due to the bias. For a charge sequence contributing to
$k_{n,\ell}^{(\pm)}$, we have the constraints $\sum_j u_j = \pm n$ and $\sum_i \upsilon_i =\pm n$. 
The term $R_{n,\ell}^{(\pm)}$ represents the sum of all $(2\ell-1)$-fold Coulomb integrals with the correlator  
(\ref{infl}) which are compatible with the constraints.

The property $Q(t -{\rm i}\,/T) = Q^\ast(t)$ in the correlator (\ref{infl})  ensures detailed balance between the 
forward and backward partial rates $k_{n,\ell}^{(\pm)}$ \cite{hbaur},
\begin{equation}
 k_{n,\ell}^{(-)} \; = \; {\rm e}^{-n \epsilon/T}_{}\,k_{n,\ell}^{(+)} \; .
\end{equation}
Actually, detailed balance already holds pairwise between particular  subsets with fixed time-order along the Keldysh contour.
The subset pairs are related by charge conjugation and time-reversal. For instance, in order $\Delta^4$
there are detailed balance relations between
(f.2) and (a.2), between (c.1) and (a.1), between (e.1) and (b.1), and between (f.1) and (d.1).
(cf. Fig.~\ref{f2} and Ref.~\cite{hbaur} for details). 

Unfortunately, the Coulomb integrals determining the various contributions to the partial rate $k_{n,\ell}^{(\pm)}$ can 
not be evaluated in analytic form for general $T$ and general $K$.  

However at $T=0$, one may use besides the detailed balance relations for the subset pairs
also scaling properties of the integrals because of the logarithmic interactions at $T=0$. These together  yield 
a multitude of relations between different Coulomb integrals of same order $\ell$ \cite{hbaur}. 
As a result of the analysis, one finds formidable cancellations in the sum of the various  Coulomb integrals of given order $\ell$, 
and only those give rate contributions at $T=0$, where the charges $\{u_j\}$ and $\{\upsilon_i\}$ are all positive, i.e., 
only direct paths without detours,
\begin{equation}\label{zero1}
k_n^{(+)} \;=\; R_{n,n}^{(+)} \,x^{2n}_{} \;,\qquad  k_n^{(-)} \;=\; 0 \;.
\end{equation}

Since $k_n^{(+)}\propto x^{2 n}$ at $T=0$ , it is straightforward to deduce from the perturbative series (\ref{tbseries}) the analytic expression
\begin{equation}\label{kzero}
 k_n^{(+)} \;=\; \frac{(-1)^{n-1}}{n!}\,\frac{\Gamma(\frac{3}{2})
\Gamma(nK)}{\Gamma(\frac{3}{2}- n + nK )} \frac{\epsilon}{2\pi}\,\left( \frac{\epsilon_0^{}}{\epsilon}\right)^{2(1-K)n}  \; .
\end{equation}
Expression (\ref{charfun}) with (\ref{kzero}) yields an analytic form of the  characteristic function. In addition, 
we get the concise moment relation
\begin{equation}
\langle I^{(m)}\rangle_{\rm c} \;=\; \left( \frac{x}{2}\,\frac{{\rm d}}{{\rm d}x}\right)^{m-1} \,\langle I \rangle \; .
\end{equation}
Thus, all statistical properties of the quantum transport process at $T=0$ can be deduced directly from the current \cite{sw01}.

Let us finally return to finite temperature.
Evidently, the analytic continuation rule in Eq.~(\ref{cond1}) can not provide information about individual tunneling rates. Rather we find with 
use of (\ref{curr}) and (\ref{rateseries}) 
\begin{equation}
{\rm Im}\,[\,\mathcal{F}(x,p) -\mathcal{F}(x,-p)\,] \;= \; \sum_{\ell\,=\,1}^\infty \sum_{n\,=\,1}^\ell \frac{n}{\ell}
\left( k_{n,\ell}^{(+)} - k_{n,\ell}^{(-)}\right) \; .
\end{equation}
This yields for the term of order $x^{2\ell}_{}$ 
\begin{equation}\label{fcontrel}
T x^{2\ell}\, {\rm Im}\,[\,\mathcal{C}_{2\ell}(-p) -\mathcal{C}_{2\ell}(p)\,]  \;=\; \sum_{n\,=\,1}^\ell 
\frac{n}{\ell}\left( k_{n,\ell}^{(+)} - k_{n,\ell}^{(-)}\right)
\end{equation}
Thus, the analytically continued perturbative series of the free energy yields for each term, say $\mathcal{F}_\ell$, a particular linear combination
of partial rates of same order $x^{2\ell}$, as given in Eq.~(\ref{fcontrel}). In order to calculate individual partial tunneling rates,
which are required in the full counting statistics at finite temperature, an analysis of the Coulomb integrals in
real-time  is indispensable. 

\section{Conclusions}
\label{sconcl}

We have discussed a variety of seemingly different physical models which display a wide range of interesting characteristics
and can be realized experimentally. These models have in common that they can be mapped on a one-dimensional 
quantum field with gapless bulk excitations and a sinusoidal  boundary interaction.  
The massless boundary sine-Gordon model represents  an impurity in a Luttinger liquid. When the Luttinger liquid modes 
are integrated out, the model describes a quantum-dissipative particle moving in a tilted sinusoidal potential. 
All these models can be successfully treated by a variety of interesting and powerful techniques.
In most cases, the task of calculating nonlinear transport and full counting statistics requires 
application of nonequilibrium techniques. For general model parameters, this is a challenging endevour. Despite the fact that the 
systems are integrable with help of the thermodynamic Bethe ansatz, this method can not be used to calculate
the full counting statistics, apart from the current. 
To gather full information about the statistical fluctuations, a real time calculation of the
individual tunneling rates is essential.

Here we pursued a different route. We seized a suggestion by Fendley, Ludwig and Saleur, who calculated the 
twisted partition function of a "log-sine" Coulomb gas on a ring, and analyzed a conjectured relation with the nonlinear conductance which involves analytic continuation in the bias and coupling strength. We studied the weak-tunneling limit, the classical regime,
and the zero temperature case and found in all cases agreement with results available from a non-equilibrium real-time calculation.

\begin{appendix}

\section{The cumulants $\mathcal{C}_4(p)$ and $\mathcal{C}_6(p)$}
\label{sapp}

\subsection{$\mathcal{C}_4(p)$}

With the substitution (\ref{asy1}) and use of the asymptotic expansion (\ref{asy}), the expressions (\ref{c412}) and turn into
\begin{eqnarray}\label{int41}
\mathcal{C}_4^{(1)}(p) &=&  \frac{1}{2}\frac{p^{4K-3}}{\Gamma^4 (K)} \int_0^\infty\!\! {\rm d}u \, f^{\,2}(u) \; ,\\  
\mathcal{C}_4^{(2)}(p) &=&  \frac{p^{4K-3}}{\Gamma^4 (K)} \int_0^\infty\!\! {\rm d}u \, f(u) h(u)\,
\int_0^{\,u}\!\! {\rm d}\upsilon \, f(\upsilon) \; ,  \label{int42}
 \end{eqnarray}
 where
\begin{eqnarray}
f(u) &=& u^{K-1}(1+u)^{K-1} \;,  \\[2mm]
h(u) &=& K(K-1)\big({\textstyle \frac{1}{u}+\frac{1}{1+u}}\big) \;.
\end{eqnarray}
The function $h(u)$ originates from the terms of order $1/p$ in the curly bracket of (\ref{asy}).
The integrals (\ref{int41}) and (\ref{int42}) are convergent in the regime $\frac{1}{2}< K <\frac{3}{4}$.
Observing that 
\begin{equation}\label{frel}
f(u)h(u) \;=\; K \,f\,'(u)  \; ,
\end{equation}
where the prime indicates differentiation with respect to the variable, the double integral
can be reduced by partial integration to a single integral, which is a Beta function. In the end, we get
\begin{eqnarray}\label{intc41}
\mathcal{C}_4^{(1)}(p) &=&  \frac{1}{2}\;\frac{p^{4K-3}}{\Gamma^4 (K)} \;\frac{\Gamma(2K-1)\Gamma(3-4K)}{\Gamma(2-2K)}  \; ,\\  
\mathcal{C}_4^{(2)}(p) &=&   - 2 K \;\mathcal{C}_4^{(1)}(p) \; .
\end{eqnarray}
This yields $a_2^{}(K) =1-2K$.
Thus, in the sum of $\mathcal{C}_4^{(1)}(p)$ and $\mathcal{C}_4^{(2)}(p)$ the singularity at $K=1/2$ is abrogated, and we end up with the cumulant expression
\begin{equation}
\mathcal{C}_4(p) \;=\; -\; \frac{1}{2}\;\frac{\Gamma(2K)}{\Gamma^4 (K)} \;\frac{\Gamma(3-4K)}{\Gamma(2-2K)}\, p^{4K-3} \; ,
\end{equation}
which is regular in the range $0< K <\frac{3}{4}$.

\subsection{$\mathcal{C}_6(p)$}

As $|p|\to \infty$, way may use (\ref{asy1}) and the asymptotic series (\ref{asy}) in the expressions (\ref{c6123}). We then obtain
\begin{eqnarray}\label{intc61}
\mathcal{C}_6^{(1)}(p) &=&  \frac{1}{3}\frac{p^{6K-5}}{\Gamma^6 (K)} \int_0^\infty\!\! {\rm d}u \, f^3(u) \; ,\\[2mm]  \label{intc62}
\mathcal{C}_6^{(2)}(p) &=& g\, \frac{p^{6K-5}}{\Gamma^6 (K)}\left\{
\int_0^\infty \!\!{\rm d}u \, \left( f^2(u)  \right)' \int_0^{\,u}\!\! {\rm d}\upsilon \,f(\upsilon) \right.    \\  \nonumber 
&&\qquad\qquad + \left.  \int_0^\infty\!\! {\rm d}u \, f\,'(u)\int_0^{\,u}\!\!{\rm d}\upsilon \,f^{\,2}(\upsilon)  \right\}  \; ,\\[2mm]  \label{intc63}
\mathcal{C}_6^{(3)}(p) &=& K^2\, \frac{p^{6K-5}}{\Gamma^6 (K)}\\  \nonumber
&&\times \;\left\{
\int_0^\infty\!\!{\rm d}u \, f\,''(u)  \int_0^{\,u}\!\!{\rm d}\upsilon \, f(\upsilon)   \int_0^{\,\upsilon} \!\!{\rm d}w \, f(w)  \right. \\ \nonumber  %\label{intc63}
&&  + \left. 2 \int_0^\infty\!\!{\rm d}u \, f\,'(u) \int_0^{\,u}\!\!{\rm d}v \, f\,'(\upsilon) 
\int_0^{\,\upsilon}\!\!{\rm d}w \, f(w) \right\}  \; .
\end{eqnarray}
Here we have again employed the relation (\ref{frel}). The integrals in (\ref{intc61}) - (\ref{intc63}) are convergent in the region
$\frac{2}{3}< K <\frac{5}{6}$. The double integrals (\ref{intc62}) and triple integrals (\ref{intc63}) may be transformed by partial integrations into single integrals, which again are representations of the Beta function.  We get
\begin{eqnarray}\nonumber
\mathcal{C}_6^{(1)}(p) &=&  \frac{1}{3}\frac{p^{6K-5}}{\Gamma^6 (K)} \,\frac{\Gamma(3K-2)\Gamma(5-6K)}{\Gamma(3-3K)} \; , \\
\mathcal{C}_6^{(2)}(p) &=& - \frac{9}{2}\,K\,\mathcal{C}_6^{(1)}(p)  \; ,  \\   \nonumber
\mathcal{C}_6^{(3)}(p) &=&  \frac{9}{2}\,K^2\,\mathcal{C}_6^{(1)}(p)  \; .
\end{eqnarray}
The sum of these terms yields $C_6(p) = a_3(K)\, C_6^{(1)}(p)$ with 
\begin{equation}\label{prefac3}
a_3^{}(K)= \frac{1}{2} (3K-1)(3K-2)\;.
\end{equation}
The prefactor (\ref{prefac3}) just cancels the singularities of $\mathcal{C}_6^{(1)}(p)$ located at $K=\frac{2}{3}$ and at $K=\frac{1}{3}$.
While $\mathcal{C}_6^{(1)}(p)$ is regular in the range $ \frac{2}{3}<K<\frac{5}{6}$, the resulting cumulant expression
\begin{equation}
\mathcal{C}_6 (p) \;=\; a_3(K)\, \mathcal{C}_6^{(1)}(p) \;=\;
\frac{1}{6}\,\frac{\Gamma(3K)}{\Gamma^6 (K)} \,\frac{\Gamma(5-6 K)}{\Gamma(3-3 K)}\,p^{6 K-5} \; ,
\end{equation}
is regular in the extended regime $0\le K < \frac{5}{6}$.

\end{appendix}

\end{document}